\documentclass[longauth]{aa}  
\usepackage{graphicx}
\usepackage{txfonts}
\usepackage{amstext}
\usepackage{color}

\def\cxou{CXOU~J171419.8-383023}

\begin{document}
   \title{Discovery of a VHE gamma-ray source coincident with the supernova remnant CTB~37A}
   \titlerunning{HESS~J1714-385}
   \authorrunning{The H.E.S.S. Collaboration}


   \author{F. Aharonian\inst{1,13}
 \and A.G.~Akhperjanian \inst{2}
 \and U.~Barres de Almeida \inst{8} \thanks{supported by CAPES Foundation, Ministry of Education of Brazil}
 \and A.R.~Bazer-Bachi \inst{3}
 \and B.~Behera \inst{14}
 \and M.~Beilicke \inst{4}
 \and W.~Benbow \inst{1}
 \and K.~Bernl\"ohr \inst{1,5}
 \and C.~Boisson \inst{6}
 \and V.~Borrel \inst{3}
% ...
}

\institute{
Max-Planck-Institut f\"ur Kernphysik, P.O. Box 103980, D 69029
Heidelberg, Germany
\and
 Yerevan Physics Institute, 2 Alikhanian Brothers St., 375036 Yerevan,
Armenia
\and
Centre d'Etude Spatiale des Rayonnements, CNRS/UPS, 9 av. du Colonel Roche, BP
4346, F-31029 Toulouse Cedex 4, France
\and
Universit\"at Hamburg, Institut f\"ur Experimentalphysik, Luruper Chaussee
149, D 22761 Hamburg, Germany
\and
Institut f\"ur Physik, Humboldt-Universit\"at zu Berlin, Newtonstr. 15,
D 12489 Berlin, Germany
\and
LUTH, Observatoire de Paris, CNRS, Universit\'e Paris Diderot, 5 Place Jules Janssen, 92190 Meudon, 
France
\and
IRFU/DSM/CEA, CE Saclay, F-91191
Gif-sur-Yvette, Cedex, France
\and
University of Durham, Department of Physics, South Road, Durham DH1 3LE,
U.K.
\and
Unit for Space Physics, North-West University, Potchefstroom 2520,
    South Africa
\and
Laboratoire Leprince-Ringuet, Ecole Polytechnique, CNRS/IN2P3,
 F-91128 Palaiseau, France
\and 
Laboratoire d'Annecy-le-Vieux de Physique des Particules, CNRS/IN2P3,
9 Chemin de Bellevue - BP 110 F-74941 Annecy-le-Vieux Cedex, France
\and
Astroparticule et Cosmologie (APC), CNRS, Universite Paris 7 Denis Diderot,
10, rue Alice Domon et Leonie Duquet, F-75205 Paris Cedex 13, France
\thanks{UMR 7164 (CNRS, Universit\'e Paris VII, CEA, Observatoire de Paris)}
\and
Dublin Institute for Advanced Studies, 5 Merrion Square, Dublin 2,
Ireland
\and
Landessternwarte, Universit\"at Heidelberg, K\"onigstuhl, D 69117 Heidelberg, Germany
% ...
}

   \offprints{A.~Fiasson (Armand.Fiasson@lpta.in2p3.fr)}

   \date{}

 
  \abstract
   {}
{ The supernova remnant (SNR) complex CTB~37 is an interesting candidate for observations with Very High Energy (VHE) $\gamma$-ray telescopes such as H.E.S.S. In this region, three SNRs are seen. One of them is potentially associated with several molecular clouds, a circumstance that can be used to probe the acceleration of hadronic cosmic rays.}
{This region was observed with the H.E.S.S. Cherenkov telescopes and the data were analyzed with standard H.E.S.S. procedures. Recent X-ray observations with Chandra and XMM-Newton were used to search for X-ray counterparts.}
{The discovery of a new VHE $\gamma$-ray source HESS~J1714-385 coincident with the remnant CTB~37A is reported. The energy spectrum is well described by a power-law with a photon index of $\Gamma =2.30\pm0.13$ and a differential flux at 1~TeV of $\Phi_0 = (8.7 \pm 1.0_{\mathrm{stat}} \pm 1.8_{\mathrm{sys}}) \times 10^{-13}\text{cm}^{-2}\text{s}^{-1}\text{TeV}^{-1}$. The integrated flux above 1~TeV is equivalent to 3\% of the flux of the Crab nebula above the same energy. This VHE $\gamma$-ray source is a counterpart candidate for the unidentified EGRET source 3EG~J1714-3857.
The observed VHE emission is consistent with the molecular gas distribution around CTB~37A; a close match is expected in a hadronic scenario for $\gamma$-ray production. The X-ray observations reveal the presence of thermal X-rays from the NE part of the SNR. In the NW part of the remnant, an extended non-thermal X-ray source, \cxou, is discovered as well. Possible connections of the X-ray emission to the newly found VHE source are discussed.
}
   {}

   \keywords{ISM: supernova remnants - Gamma rays: observations - X-rays: individuals: G348.5+0.1}

   \maketitle

\section{Introduction}
The origin of Galactic cosmic rays (CRs) has been an open question ever since their discovery in 1912 by Victor Hess. It is commonly believed that supernova remnants (SNRs), or more precisely the strong shocks associated with them, are the main particle accelerators in the Galaxy up to 10$^{15}$~eV (for a recent review see e.g. Hillas 2005). A conversion efficiency of 10\% of the kinetic energy of the Galactic SNRs into CRs can explain the observed flux at Earth (taken to be typical of the Galaxy). Recently, VHE ($E>$100~GeV) $\gamma-$ray emission has been detected from several shell-type SNRs with H.E.S.S., which confirms that these objects accelerate particles up to at least 100~TeV (Aharonian et al. 2007a, Aharonian et al. 2007b). Various processes can produce VHE $\gamma$-rays, such as inverse Compton scattering by accelerated electrons, or neutral pion decay ($\pi^{0}\rightarrow\gamma\gamma$) after hadronic interactions of accelerated protons. The observed $\gamma$-ray emission in a narrow energy band is not sufficient to disentangle the contributions from these processes. Hadronic interactions require, however, a significant amount of target matter to produce a detectable $\gamma$-ray flux. The observation of dense molecular clouds in the vicinity of supernova blast waves could thus be a probe of proton acceleration by supernova remnants (Aharonian et al. 1994).

Given the uncertainties in estimating the distances of molecular clouds and SNRs, a directional coincidence is only a necessary condition for a physical association of these objects. The additional presence of OH masers (1720~MHz) indicates such associations as these masers occur in shocked molecular clouds (Elitzur 1976, Frail et al. 1996). In this context, the region of the SNR complex CTB~37 is particularly interesting for observations with VHE $\gamma$-ray instruments. Three young SNRs are seen in this region and one of these remnants is interacting with several molecular clouds. CO (J=1$\rightarrow$0) emission as well as OH maser emission at 1720~MHz has been detected at various locations towards the SNR~G348.5+0.1 (also labeled as CTB~37A; Frail et al. 1996).

Gamma-ray emission from such an interaction between a SNR and a molecular cloud may have already been detected by VHE telescopes. The northern $\gamma$-ray source in the W28 region, HESS~J1801-233, is coincident with molecular clouds overtaken by the forward shock of the remnant as revealed by OH masers (Aharonian et al. 2008a). In the direction of IC443, the discovery of a VHE $\gamma$-ray excess with MAGIC has been reported (Albert et al. 2007). Also this emission is coincident with a molecular cloud and OH masers. W28 and IC443 both belong to the mixed-morphology SNR class, which appears to be linked to the interaction with dense molecular clouds (Yusef-Zadeh et al. 2003).

In this paper, the nature of the VHE emission from HESS~J1714-385 coincident with G348.5+0.1 will be discussed. While hadronic high-energy particles primarily lead to emission in the $\gamma$-ray band, a population of multi-TeV electrons that emit VHE $\gamma$-rays should be accompanied by synchrotron X-ray emission of these electrons in the ambient magnetic field. A population of protons with energies higher than a few 10~TeV should only produce much fainter X-ray emission through secondary electrons produced in hadronic interactions. An analysis of recent XMM-Newton and Chandra X-ray data has been made to search for X-ray counterparts.

\section{H.E.S.S. observations and results}
H.E.S.S. (High Energy Stereoscopic System) is an array of four imaging atmospheric Cherenkov telescopes located 1800 meters above sea level, in the Khomas Highland of Namibia (Bernl{\"o}hr et al. 2003). Each 13~m diameter telescope is located at a corner of a 120~m square and is equipped with a camera composed of 960~photomultiplier tube pixels (Vincent et al. 2003). Each pixel has a field of view of 0.16~degrees, leading to a total field for the camera of 5~degrees. Gamma-ray events can be reconstructed with an angular resolution of $\sim$0.1~degrees. The sensitivity for a point-like source reaches $\sim$1\% of the Crab nebula flux for a 25~h exposure.

An analysis of H.E.S.S. observations of the region of the SNR complex CTB~37 has already been published, in the context of the H.E.S.S. Galactic plane survey (Aharonian et al.~2006a). A source of VHE $\gamma$-rays was detected, HESS~J1713-381, coincident with the supernova remnant G348.7+0.3 (CTB~37B). This analysis also showed a second excess coincident with the SNR~G348.5+0.1, but the signal was not significant at that time.

\begin{figure}
\begin{center}
\includegraphics [width=0.45\textwidth]{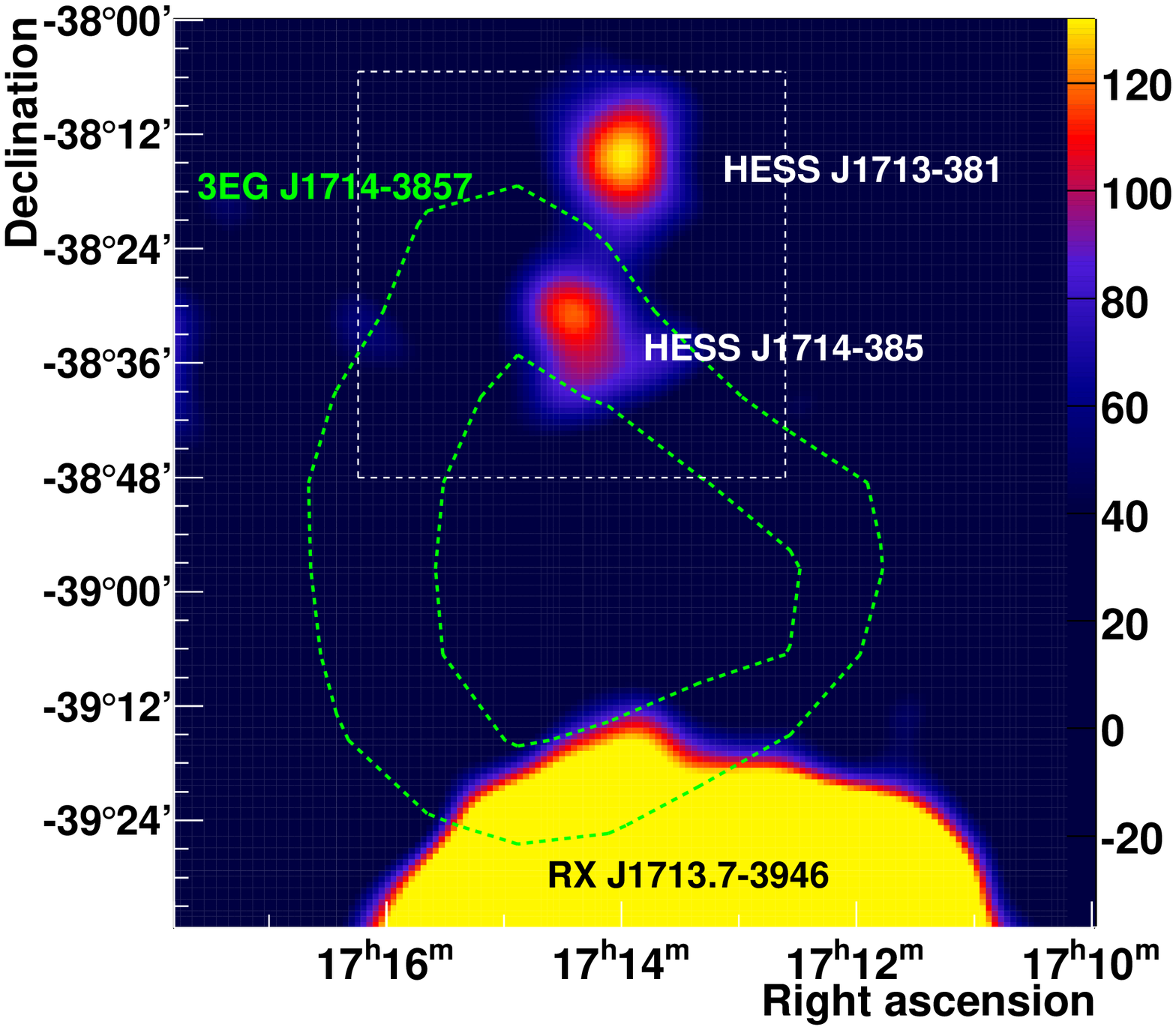}
\includegraphics [width=0.45\textwidth]{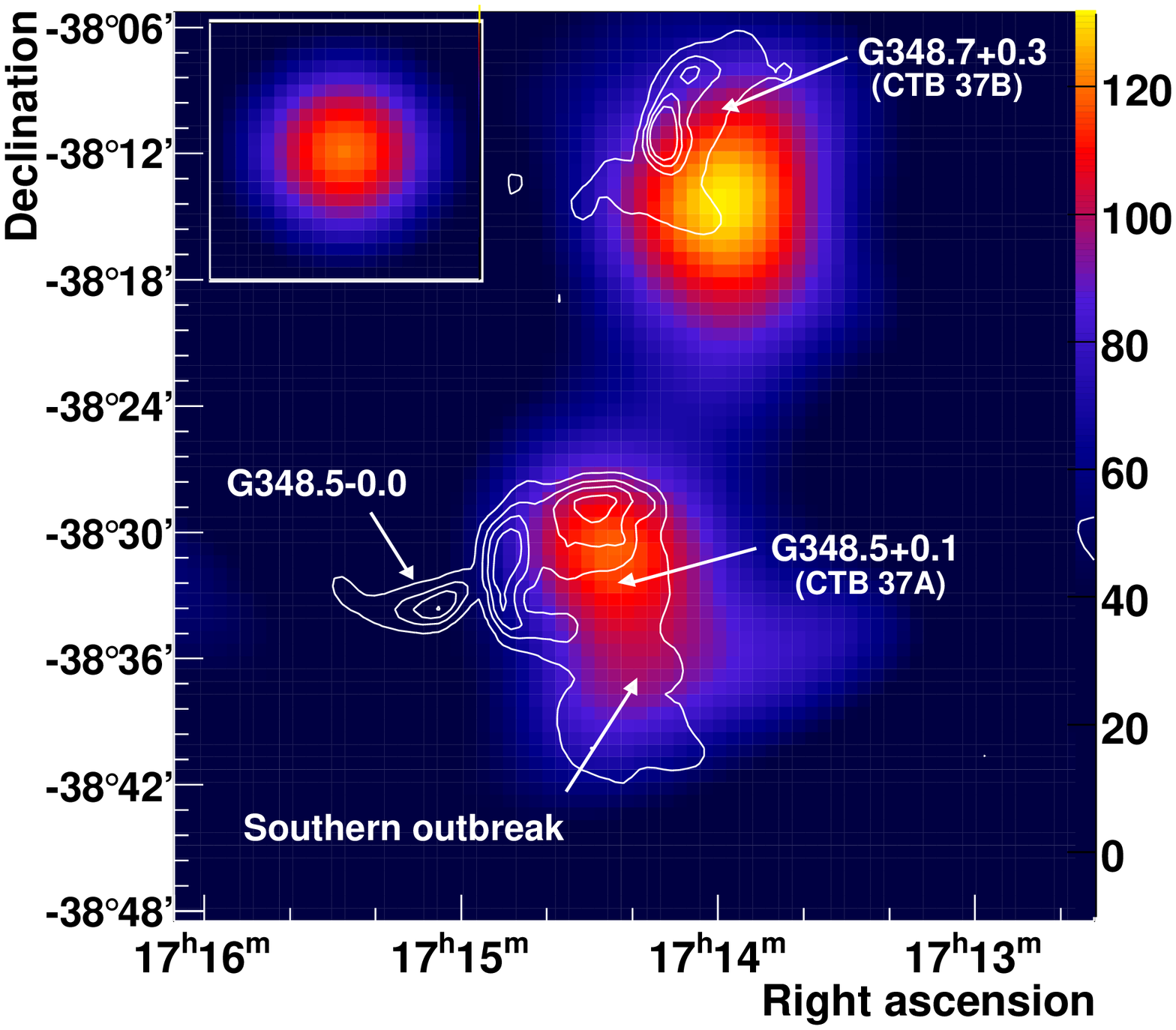}
\end{center}
\caption{\textit{Top:} H.E.S.S. excess map of the SNR complex CTB~37 region obtained with cuts B. The map is smoothed with a $\sigma_{\mathrm{sm}}=2.9'$ Gaussian. The color scale is in units of counts per 2$\pi\sigma_{\mathrm{sm}}^2$. The 95\% and 68\% confidence levels of the position of 3EG~J1714-3857 are overlaid as green contours. The color scale has been saturated in order to increase the visibility of HESS~J1713-381 and HESS~J1714-385. \textit{Bottom:} An expanded view of the top white dotted box. The white contours are radio (843~MHz) 0.1, 0.5, 0.9 \& 1.4~Jy/beam contours from the Molonglo Galactic Plane Survey (Green et al.~1999). The PSF of the instrument, smoothed in the same way as the excess map, is represented in the inset panel.}
\label{fig1}
\end{figure}

More observations have been taken in this region since then. The SNR complex CTB~37 is in the field of view of most of the observations taken around RX~J1713.7-3946, and benefits from a deep exposure. The current dataset includes all runs within 2 degrees distance between the SNR~G348.5+0.1 and the centre of field of view position. After data quality selection and dead-time correction, the resulting live time is 67.6~hours (equivalent to 42.7~hours of on-axis exposure). The observations have been performed in a large range of zenith angles, from 14 to 71~degrees, with an average value of 39~degrees.

These data have been analyzed using a combined Model-Hillas analysis (de Naurois~2006). This method consists of a comparison of shower images with a semi-analytical model, combined with Hillas parameters estimation. Event selection is made based on a combined estimator (Combined Cut 2) and shower image properties. The cuts used for this analysis are optimized for searches of faint sources (Table~\ref{cut_list}). Background estimation has been performed using the ring background method for sky maps, and the reflected-region technique for spectral analysis (for more details see Berge et al. 2007). The energy threshold of this analysis is 200~GeV with cuts A applied for spectral extraction, and  310~GeV with cuts B applied for sky maps, source search, and source position determination. An independent standard H.E.S.S. analysis using the Hillas moment-analysis scheme (Aharonian et al. 2006b) and separate calibration scheme has also been made. The results of both analyses agree within errors. The results obtained with cuts A and cuts B are also consistent.

\begin{table}[h]
\begin{center}
\begin{tabular}{lcc}
\hline
 &    cuts A &  cuts B \\

\hline
\hline
  Combined cut 2 max. & 0.7 & 0.7 \\
  Shower depth min. (rad. length) & -1 & -1 \\
  Shower depth max. (rad. length) & 4 & 4 \\
  Image size min. (photo-electrons)& 60  & 60 \\
  Nominal distance max. (degrees) & 2.5 & 2.5\\
  Event multiplicity (telescopes) & $\geq$~2 & $\geq$~3\\
\hline
\end{tabular}
\end{center}
\caption{List of cuts used in this weak source analysis. The shower depth is the shower reconstructed primary interaction depth. The nominal distance is the distance of the image barycenter to the center of the camera. The event multiplicity is the number of images satisfying the size and nominal distance requirements.}
\label{cut_list}
\end{table}

A search for a point-like emission has been performed in this region within the updated dataset. Figure~\ref{fig1} shows the resulting excess map. The excess previously reported, close to HESS~J1713-381 and coincident with G348.5+0.1, is confirmed with a peak significance of 10.1~$\sigma$ (using an integration radius of $7.8'$). The number of such test positions within the H.E.S.S. Galactic scan is estimated to $\approx6.5\times10^5$ (Aharonian et al. 2006a). Accounting for this number of trials, the statistical significance of the signal is 8.7~$\sigma$ post-trials.

\begin{figure}
\begin{center}
\includegraphics [width=0.5\textwidth]{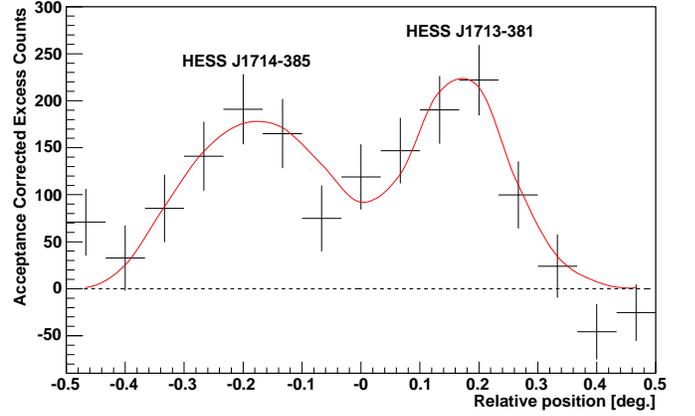}
\end{center}
\caption{Emission profile along the axis defined by the two sources HESS~J1714-385 and HESS~J1713-381. A profile of the best fit model is shown as a red solid curve.
}
\label{fig2}
\end{figure}

A joint fit of HESS~J1713-381 and the new observed excess has been made using two Gaussians convolved with the instrument point spread function (PSF). The residual emission observed between both sources does not constitute an excess beyond the anticipated contribution from each source at this location and supports the idea that the two sources are independent (Figure~\ref{fig2}). The discovery of a new VHE $\gamma$-ray source is thus announced, to which is assigned the identifier HESS~J1714-385. The new data set also allows a deeper study of HESS~J1713-381, the results of which are presented in a separate paper (Aharonian et al.~2008b).

From the fit, the position of HESS~J1714-385 has been extracted: 17${^{\mbox{\tiny h}}}$14${^{\mbox{\tiny m}}}$19${^{\mbox{\tiny s}}}$ , $-38^{\circ}34'$ (J2000) with $1'20''$ statistical error in Right Ascension and Declination. The source extension is of the same order as the analysis PSF, therefore no morphological information can be extracted beyond the fact that the source is extended with an RMS size of 4$'$ $\pm$ 1$'$ (cf. PSF 68\% containment radius of 0.072~degrees). A fit with an asymmetrical Gaussian does not improve the fit quality.

\begin{figure}
\begin{center}
\includegraphics [width=0.5\textwidth]{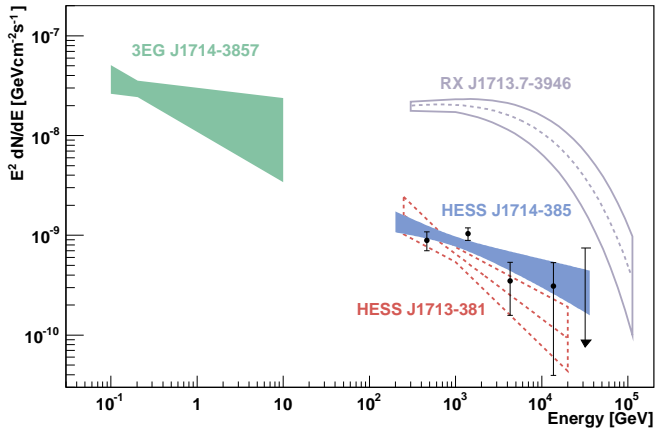}
\end{center}
\caption{Reconstructed energy spectrum of HESS~J1714-385 (blue) in the energy range between 200~GeV and 40~TeV, in comparison to RX~J1713.7-3946 (purple; Aharonian et al. 2007a), HESS~J1713-381 (red; Aharonian et al. 2008b), and 3EG~J1714-3857 (green; Hartman et al. 1999). The differential flux points shown are computed by multiplying the fractional residuals between detected and modeled photon-count spectrum with the modeled flux spectrum.}
\label{fig3}
\end{figure}

The energy spectrum has been derived using a forward folding method (Piron et al. 2001) in an integration region of radius 0.2~degrees centered on the fitted position. There are 975 excess $\gamma$-events within this region, after applying cuts A. Using the fitted Gaussians, the contamination from HESS~J1713-381 in the integration circle of HESS~J1714-385 is estimated to be 5\%. The reconstructed spectrum, in the energy range between 200~GeV and 40~TeV, is compatible with a power-law of the form $dN/dE\,=\,\Phi_0(E/\mathrm{1 TeV})^{-\Gamma}$ with a spectral index $\Gamma=2.30\pm0.13_{\mathrm{stat}}\pm0.20_{\mathrm{sys}}$ and a differential normalisation at 1~TeV of $\Phi_0 = (8.7 \pm 1.0_{\mathrm{stat}} \pm 1.8_{\mathrm{sys}}) \times 10^{-13}\text{cm}^{-2}\text{s}^{-1}\text{TeV}^{-1}$ ($\chi^2$/dof = 27.4/24). The integrated flux above 1~TeV corresponds to 3\% of the Crab nebula flux above the same energy (Aharonian et al. 2006b). Figure~\ref{fig3} shows the reconstructed VHE $\gamma$-ray spectrum.

There is no indication for variability of the flux. Given the fact that the source is close to the sensitivity threshold, and that the data are taken within only 4 months every year, the data was binned roughly monthly. The $\chi^2$ per degree of freedom of a constant fit to the resulting light curve is 7.3/7.

\section{Association with SNR G348.5+0.1 and molecular clouds}

The SNR G348.5+0.1 (also called CTB~37A) is part of an unusual SNR complex composed of three SNRs (Figure~\ref{fig1}~{\it bottom}); the radio source CTB~37A was originally believed to be one single SNR but is now thought to consist of two remnants, G348.5+0.1 (still called CTB~37A) and G348.5-0.0 (Kassim et al. 1991). The remnant G348.5+0.1 is well defined in its Northern part and appears more extended to the South. This break-out morphology might be due to expansion into an inhomogeneous medium. A constraint on the distance is obtained from neutral hydrogen absorption and gives 10.3~$\pm$~3.5 kpc (Caswell et al. 1975). Clark \& Stephenson (1977) proposed either G348.5+0.1 or G348.7+0.3 as candidates for the remnant of the SN of \footnotesize AD \normalsize 393. Downes (1984) remarks that the high surface brightness of G348.5+0.1 would be consistent with this hypothesis. The extension of the shell, $9.5'\times8'$ semi-major and semi-minor axes including the outbreak (Whiteoak \& Green 1996), is compatible with the $4'$ extension of the VHE $\gamma$-ray emission (Gaussian width), following the argument of Aharonian et al. (2008b). Hence, from size and position arguments, an association with the whole shell cannot be excluded. However, as argued in the following, the molecular clouds associated with G348.5+0.1 also prove to be a plausible counterpart to the VHE source.

Several OH masers at 1720~MHz have been detected towards SNR G348.5+0.1 (Frail et al. 1996). They are distributed in the interior and along the edge of the SNR. Most of the masers have a velocity close to -65~km~s$^{-1}$ and are coincident with three molecular clouds observed in the CO (J=1$\rightarrow$0) transition (115~GHz) at the same velocity (Reynoso \& Mangum 2000). These authors estimate the distance of the clouds to be 11.3 kpc, which implies a size of the remnant close to 28~pc. The velocity of the masers as well as their superposition with the clouds argue in favor of their physical association with the molecular clouds. This provides a strong indication that molecular clouds are being overtaken by the forward shock of G348.5+0.1.

\begin{figure*}
\begin{center}
\includegraphics [width=0.7\textwidth]{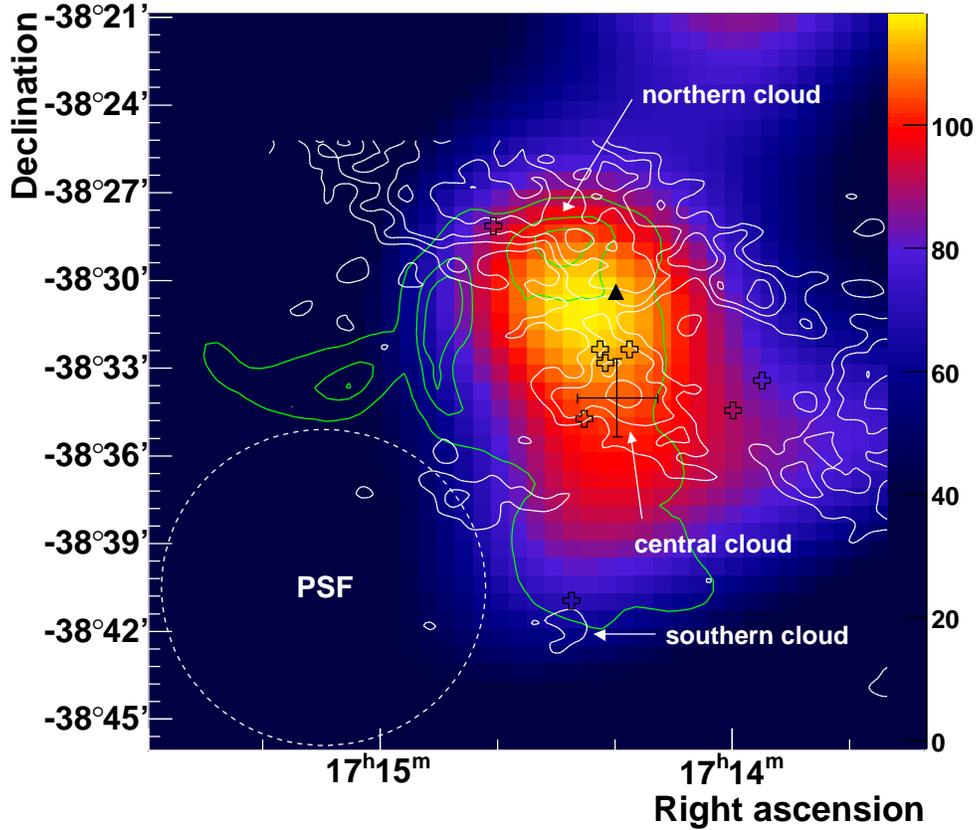}
\end{center}
\caption{
An expanded view of the VHE $\gamma$-ray excess map around HESS~J1714-385 (the color scale is in unit of counts per 2$\pi\sigma_{\mathrm{sm}}^2$). The map is smoothed with a $\sigma_{\mathrm{sm}}=2.9'$ Gaussian. 
The 0.1, 0.9 \& 1.4~Jy/beam radio contours (843~MHz) from the Molonglo Galactic Plane Survey (Green et al.~1999) are overlaid in green. The white contours are CO emission at 17, 25, 33, 41 and 49 K~km~s$^{-1}$, integrated between -68~km~s$^{-1}$ and -60~km~s$^{-1}$, from the NRAO 12~m Telescope (Reynoso \& Mangum 2000). The contours are truncated toward the North because the CO data only extends up to this Declination. The positions of OH masers at 1720 MHz with velocity close to -65 km$\,$s$^{-1}$ are marked with small black open crosses. The best fit position for HESS J1714-385, derived under the assumption of an azimuthally symmetric Gaussian source shape, is reported with a large black cross.
The white dashed circle illustrates the 68\% containment radius of the H.E.S.S. smoothed PSF. The black triangle indicates the position of the X-ray source \cxou.
}
\label{fig4}
\end{figure*}

Figure~\ref{fig4} shows the matter distribution surrounding the remnant, through the CO (J=1$\rightarrow$0) transition intensity integrated between -68~km~s$^{-1}$ and -60~km~s$^{-1}$, overlaid on the VHE $\gamma$-ray excess map. The position of the OH masers with a velocity close to -65~km~$^{-1}$ is also indicated. The masses of individual clouds (the northern, central, and southern cloud, as identified by Reynoso \& Mangum (2000)) range between $1.3\times10^{3}$~M$_\odot$ and $5.8\times10^{4}$~M$_\odot$ with H$_2$ densities between 150~cm$^{-3}$ and 660~cm$^{-3}$. The VHE $\gamma$-ray centroid is in good coincidence with the central cloud (see Fig.4) and the associated OH masers. This cloud is an interesting candidate for the origin of the VHE $\gamma$-ray emission. However, the extension of the VHE emission ($\sim4'$) is larger than the core extension of this cloud ($\sim1'$). The presence of OH masers towards the northern and southern clouds indicates that they are also interacting with the remnant and may therefore contribute to the VHE $\gamma$-ray emission.

Apart from the presence of masers coincident with the central cloud, another observation seems to indicate an interaction of the remnant with this cloud. An additional structure at -88 km s$^{-1}$ is present in the CO profile towards the direction of the masers (Reynoso \& Mangum 2000). This velocity could be explained by the acceleration of a fraction of the cloud by the forward shock. In this case the velocity of the shock in the molecular cloud can be constrained to $\sim$15-30 km s$^{-1}$.

Zeeman splitting of the maser emission has been studied (Brogan et al. 2000) and reveals a complex magnetic field morphology with values ranging from 0.22~mG to 1.5~mG. The values obtained for the masers coincident with the central cloud spread between 0.22~mG and 0.6~mG. 

\section{Association with the EGRET source 3EG~J1714-3857}

In the sky region of the CTB 37 complex and RX~J1713.7-3946, the EGRET source 3EG~J1714-3857 was also detected (Hartman et al. 1999; see Figure~\ref{fig1}). The new VHE $\gamma$-ray source HESS~J1714-385 is located within the 95\% confidence contour  of the EGRET source position and close to the 68\% confidence contour. The EGRET source, flagged as possibly extended or multiple, is currently unidentified. Its origin has been of particular interest as it overlaps the northern part of the SNR~RX~J1713.7-3946. Two dense molecular clouds in the surroundings of this remnant have been suggested as the source of the GeV $\gamma$-ray emission (Butt et al. 2001). However, the detection of a VHE $\gamma$-ray source close to the 68\% confidence level contour of the GeV emission and potentially associated with several molecular clouds makes the new H.E.S.S. source a good counterpart candidate. The integrated flux of the EGRET source above 100 MeV is $(46.3\pm6.5)\times10^{-8}\mbox{m}^{-2}\mbox{s}^{-1}$ with a spectral photon index of 2.3$\pm$0.2. The spectra of 3EG~J1714-3857 and HESS~J1714-385 are represented in Figure~\ref{fig3}, together with the other H.E.S.S. sources which could be associated with the EGRET source, RX~J1713.7-3946 and HESS~J1713-381 (CTB~37B), both represented as open error bands. The EGRET source spreads over a much larger region than HESS~J1714-385, and all three H.E.S.S. sources could contribute to the EGRET source. However, a global fit to the whole GeV emission and HESS~J1714-385 gives a spectral index of 2.45, quite close to the two individual spectra. Although such a compatibility between spectra is expected by chance (Funk et al. 2007), this good agreement makes the new H.E.S.S. source a good counterpart candidate for the EGRET source.

\section{X-ray observations}

The Chandra X-ray Observatory observed the remnant G348.5+0.1 on October 10, 2006 for 20~ks (ObsID 6721) and XMM-Newton observed this region on March 1, 2006 for 17~ks (ObsID 0306510101). Data have been analyzed using the Chandra Interactive Analysis of Observations (CIAO version 3.4, CALDB version 3.4.1) and the XMM-Newton Science Analysis Software (SAS version 7.1), respectively. The datasets have been cleaned from soft proton flares and the resulting observation times are 19.9~ks (not affected by flares) and 9.6~ks, respectively.

\begin{figure*}
\begin{center}
\includegraphics [width=1.\textwidth]{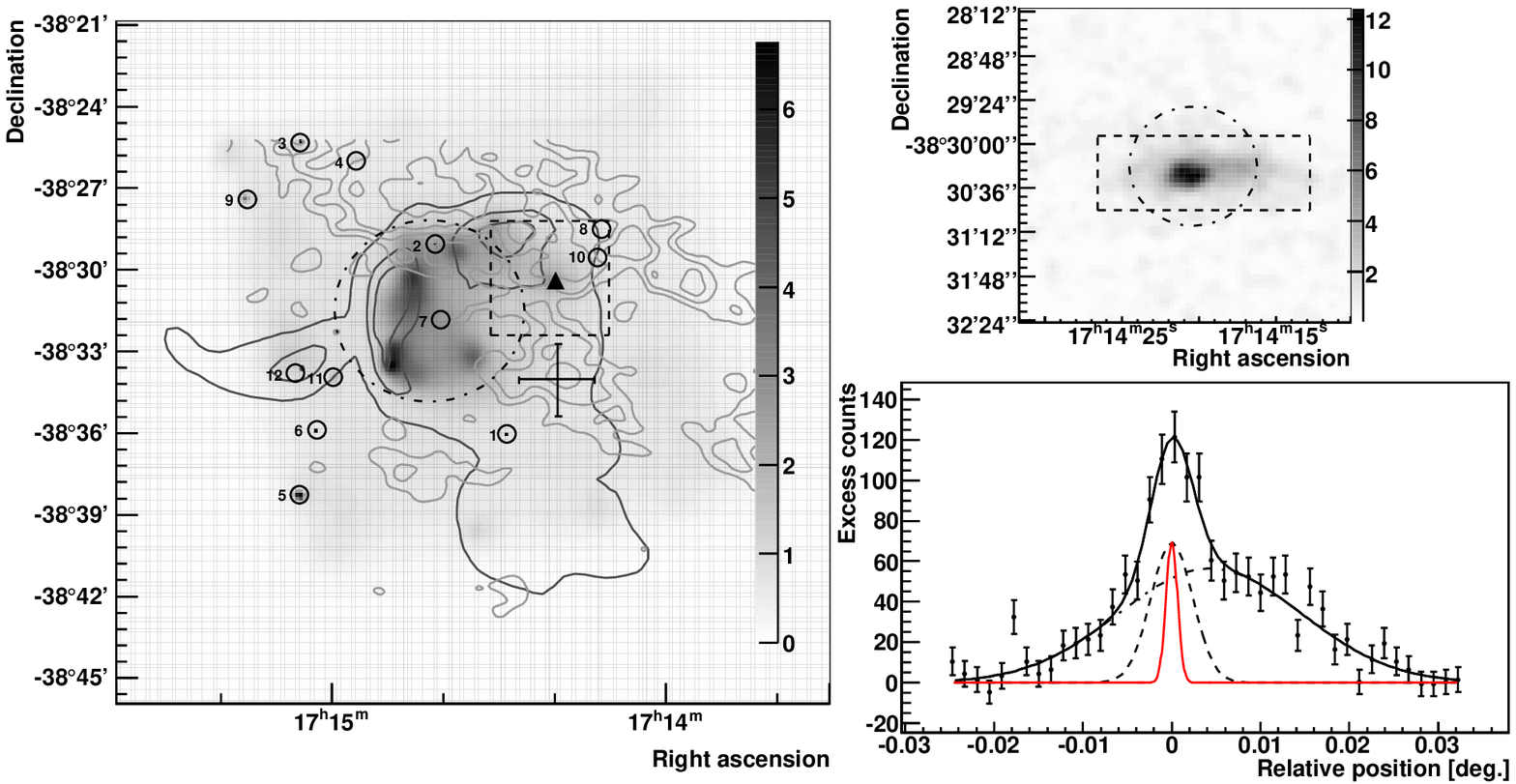}
\end{center}
\caption{
{\it Left:}~Adaptively smoothed count map from Chandra in the energy band 1.2 to 2.5~keV. The color scale has been truncated to 6.8. 0.1, 0.9 \& 1.4~Jy/beam radio contours (843~MHz) from the Molonglo Galactic Plane Survey (Green et al.~1999) are overlaid in black. The grey contours are CO emission at 17, 25, 33, 41 and 49 K~km~s$^{-1}$, integrated between -68~km~s$^{-1}$ and -60~km~s$^{-1}$, from the NRAO 12~m Telescope (Reynoso \& Mangum 2000). The contours are truncated toward the North because the CO data only extends up to this Declination. The best fit position for HESS J1714-385, derived under the assumption of an azimuthally symmetric Gaussian source shape, is reported with a large black cross. The positions of the X-ray sources detected using the CIAO {\it wavdetect} algorithm are shown as small circles. The dashed-dotted circle is the region used for the spectral analysis of the diffuse emission. The black triangle indicates the position of the X-ray source \cxou.
{\it Right top:}~An expanded view of the region indicated by the dashed box in the left panel, in the energy band 3-7 keV, after a 4$''$ Gaussian smoothing (the color scale is in unit of counts $(16\,\mathrm{arcsec})^{-2}$). The dashed-dotted circle is the extraction region used for the spectral analysis. {\it Right bottom:}~Emission profile (unsmoothed) taken from the rectangular region shown in the top right panel, centered on \cxou. The dashed and dotted-dashed lines are the two fitted Gaussian functions. The sum of these functions is represented as a solid black line. The red line is a Gaussian fit of the projected Chandra PSF at the core position.
}
\label{fig5}
\end{figure*}

\begin{table*}
\begin{center}
\begin{tabular}{cccccc}
\hline
ID & Name & RA & Dec & Counts & Significance ($\sigma$)\\
\hline
\hline
1 & CXOU~J171428.5-383601  &  17$^{\mathrm{h}}$14$^{\mathrm{m}}$28.59$^{\mathrm{s}}$ &  -38$^{\mathrm{d}}$36$'$1.6$''$& 74.6 & 35.8 \\
2 & CXOU~J171441.4-382903  &  17$^{\mathrm{h}}$14$^{\mathrm{m}}$41.4$^{\mathrm{s}}$ &  -38$^{\mathrm{d}}$29$'$3.5$''$& 37.2 & 16.5 \\
3  &  CXOU~J171505.7-382519  &  17$^{\mathrm{h}}$15$^{\mathrm{m}}$5.75$^{\mathrm{s}}$ &  -38$^{\mathrm{d}}$25$'$19.8$''$& 52.8 & 16.3 \\
4  &  CXOU~J171455.6-382559  &  17$^{\mathrm{h}}$14$^{\mathrm{m}}$55.67$^{\mathrm{s}}$ &  -38$^{\mathrm{d}}$25$'$59.7$''$& 37.3 & 12.0 \\
5  &  CXOU~J171559.0-383816  &  17$^{\mathrm{h}}$15$^{\mathrm{m}}$5.92$^{\mathrm{s}}$ &  -38$^{\mathrm{d}}$38$'$16.3$''$& 36.5 & 11.9 \\
6  &  CXOU~J171527.0-383553  &  17$^{\mathrm{h}}$15$^{\mathrm{m}}$2.76$^{\mathrm{s}}$ &  -38$^{\mathrm{d}}$35$'$53.3$''$& 35.3 & 10.6 \\
7  &  CXOU~J171440.4-383150  &  17$^{\mathrm{h}}$14$^{\mathrm{m}}$40.49$^{\mathrm{s}}$ &  -38$^{\mathrm{d}}$31$'$50.3$''$& 19.4 & 9.0 \\
8  &  CXOU~J171411.6-382831  &  17$^{\mathrm{h}}$14$^{\mathrm{m}}$11.6$^{\mathrm{s}}$ &  -38$^{\mathrm{d}}$28$'$31$''$& 20.0 & 7.5 \\
9  &  CXOU~J171515.2-382724  &  17$^{\mathrm{h}}$15$^{\mathrm{m}}$15.28$^{\mathrm{s}}$ &  -38$^{\mathrm{d}}$27$'$24.7$''$& 18.8 & 6.9 \\
10 &  CXOU~J171412.3-382932  &  17$^{\mathrm{h}}$14$^{\mathrm{m}}$12.3$^{\mathrm{s}}$ &  -38$^{\mathrm{d}}$29$'$32.8$''$& 12.8 & 6.4 \\
11 &  CXOU~J171459.8-383356  &  17$^{\mathrm{h}}$14$^{\mathrm{m}}$59.81$^{\mathrm{s}}$ &  -38$^{\mathrm{d}}$33$'$56.1$''$& 16.2 & 5.6 \\
12 &  CXOU~J171566.0-383348  &  17$^{\mathrm{h}}$15$^{\mathrm{m}}$6.63$^{\mathrm{s}}$ &  -38$^{\mathrm{d}}$33$'$48$''$& 17.2 & 5.5 \\
\hline
\end{tabular}
\end{center}
\caption{ 
Properties of the X-ray sources detected above a 5$\sigma$ level using the CIAO {\it wavdetect} algorithm. The statistical errors, both in Right Ascension and Declination, are below 1$''$ for each source.
}
\label{tab2}
\end{table*}

The count maps obtained from the two observations are very similar. Figure~\ref{fig5} {\it left} shows the adaptively smoothed count map from Chandra (1.2-2.5~keV). The position of 12 sources detected above a level of 5$\sigma$ with the CIAO {\it wavdetect} algorithm, described in Table~\ref{tab2}, are represented on the map. A region of extended emission is visible on the Eastern part of the remnant well defined in radio emission. This source shows an excess of $4214\,\pm\,94$ counts. Figure~\ref{fig5} {\it top right} is an expanded view of the adaptively smoothed count map from Chandra (3-7~keV). A more compact source, \cxou, is detected in the Western part of the remnant (using the CIAO {\it vtpdetect} algorithm). This source is also extended ($\sim1.2''\times0.5''$ of Gaussian width) compared to the PSF of the instrument at this position and presents a core-tail structure, elongated along the East-West axis. This source shows an excess of $1429\,\pm\,50$ counts.

\subsection{The extended emission}

The energy spectrum of the Eastern emission region has been derived within a circular region shown on Figure~\ref{fig5}, in the 0.5~keV~-~7~keV energy range. The fits from Chandra ACIS and the PN, MOS1 and MOS2 XMM-Newton detectors give consistent results and are compatible with absorbed thermal emission (photo-electric absorption WABS $\times$ VMEKAL model; Morisson \& McCammon 1983, Kaastra \& Mewe 1993). The errors quoted are 68\% confidence level. The fit from Chandra ACIS gives a temperature of $0.81\pm0.04$~keV and a column density of $N_{\mathrm{H}} = 3.15^{+0.13}_{-0.12} \times 10^{22}$~cm$^{-2}$ ($\chi^2$/dof = 139.4/113). A fit with an absorbed power-law is rejected, with a probability of 10$^{-13}$ ($\chi^2$/dof = 268.5/118).

This thermal emission, located within the volume well defined by the radio rim, could be explained by a physical scenario which is used to explain mixed-morphology (MM) SNRs (Rho \& Petre 1998). Thermal X-ray emission is thought to radiate from swept-up interstellar material within such remnants. Although G348.5+0.1 does not appear as a classical MM SNR, the observed X-ray morphology could be explained by the inhomogeneous medium surrounding the remnant, responsible also for the break-out radio morphology.

\subsection{\cxou}

The energy spectrum of \cxou\ has been derived in a region of 50$''$ radius centered on the excess (J2000 $17^{\mbox{\tiny h}}14^{\mbox{\tiny m}}20^{\mbox{\tiny s}}$, $-38^{\circ}30'20''$).
It is well described by an absorbed power-law both in the XMM-Newton and Chandra measurements in the energy range 0.5~keV~-~10~keV. A global fit of XMM-Newton MOS1, MOS2, PN and Chandra ACIS gives a spectral photon index of $1.32^{+0.39}_{-0.35}$, an unabsorbed energy flux between 0.5 and 10~keV of $(4.1^{+4.1}_{-2.0})\times10^{-12}$~erg$\,$cm$^{-2}$s$^{-1}$, and a column density $N_{\mathrm{H}}$~=~$5.9^{+1.8}_{-1.4} \times 10^{22}$~cm$^{-2}$ ($\chi^2$/dof = 64.6/61). A purely thermal fit results in an unrealistic temperature $\gg$~10~keV and typical SNR plasma temperatures of below a few keV are excluded ($\chi^2$/dof = 190/62 with a temperature fixed at 1~keV). 

This extended non-thermal emission could be a signature of a pulsar wind nebula (PWN), although the energy spectrum seems harder than typical for such objects (Kargaltsev et al. 2007). The spectral fit suggests larger absorption at the location of \cxou\ than at the location of the thermal emission feature, the association of which with the SNR is more obvious from its morphology.
Such a difference in column density would not rule out an association of the X-ray PWN candidate with the remnant G348.5+0.1, since the presence of molecular clouds in the direction of the remnant could explain this difference.
If the PWN interpretation holds, G348.5+0.1 would be a composite SNR. However, there is no sign of a point-like source within the non-thermal X-ray emission and there is no pulsar reported in this region at other wavelengths. There is also no obvious radio counterpart associated with the non-thermal X-ray source.

In principle, the presence of a possible PWN allows another hypothesis to explain the extended emission observed toward the Eastern part of the remnant. This emission could be produced by plasma heated by the pulsar jet as observed in direction of the PWN powered by PSR~B1509-58 (Yatsu et al. 2005). Further evaluation of this possibility is beyond the scope of this paper.

An interesting point is the non-detection of non-thermal X-rays at the location of the TeV peak. An upper limit on the flux coming from the central cloud was derived in a 1$'$ radius region, centered on $17^{\mathrm{h}}14^{\mathrm{m}}20.5^{\text{s}}$, $-38^{\circ}32'30''$ (J2000; Reynoso \& Mangum 2000). Assuming an E$^{-2}$ spectrum and the same column density as derived from the thermal emission, an upper limit (at 99\% confidence level) on the unabsorbed energy flux between 1 and 10~keV of $3.5\times10^{-13}$~erg$\,$cm$^{-2}\,$s$^{-1}$ was derived. At a distance of 11.3~kpc, this results in a luminosity upper limit of $5.3\times10^{33}$~erg$\,$s$^{-1}$ in the same energy range. 
Assuming that the absorption column derived from the PWN candidate is better representing the column density from regions inside the molecular clouds, an upper limit on the luminosity of $7.1\times10^{33}$~erg$\,$s$^{-1}$ was derived. This energy flux is a factor of $\sim$5 lower than the VHE $\gamma$-ray flux of $2.3\times10^{-12}$~erg$\,$cm$^{-2}\,$s$^{-1}$ between 1 and 10~TeV.

\section{Nature of the VHE emission}

Several mechanisms can lead to VHE $\gamma$-ray emission, depending on the nature of the accelerated particles and the interaction targets. The likelihood of the emission of VHE $\gamma$-rays from HESS~J1714-385 being caused by populations of either accelerated protons or electrons is discussed in this section. 

\subsection{Hadronic scenario}

The study of molecular clouds in the vicinity of supernova blast waves has been suggested as a promising direct probe of accelerated CRs (Aharonian et al. 1994). The presence of shocked molecular clouds coincident with the VHE $\gamma$-ray emission supports such a scenario. First, an association with the central cloud is discussed here. As this cloud seems to be shocked, at least part of the cloud should be filled with CRs driven by the blast wave. Given the fact that the target mass is known, the density of CRs needed to produce the observed VHE $\gamma$-ray flux can be estimated. The cloud dimension being small compared to the size of the remnant, a uniform CR density  throughout the entire cloud can reasonably be assumed.

The energy range of the H.E.S.S. observation (200~GeV - 40~TeV) corresponds to a CR energy range of a few~TeV to a few 100~TeV, assuming a delta-function approximation where a mean fraction $\kappa_{\pi} =0.17$ of kinetic energy is transferred to the secondary $\pi^0$ particles. The energy density $w_{\mathrm{TeV}}$ of CRs required to provide the observed VHE $\gamma$-ray emission can be estimated:  $w_{\mathrm{TeV}}\approx t_{\mathrm{pp}\rightarrow\pi^{0}}\times L_{\gamma}$(0.2~-~40~TeV)$\times (V/\mathrm{cm}^3)^{-1}\:\mathrm{eV}\,\mathrm{cm}^{-3}$, where $t_{\mathrm{pp}\rightarrow\pi^{0}}\approx 4.5\times10^{15}(n/\mathrm{cm}^{-3})^{-1}$s is the characteristic cooling time of protons through the $\pi^{0}$ production channel, $L_{\gamma}$(0.2~-~40~TeV) is the $\gamma$-ray luminosity in the H.E.S.S. energy range, and $V$ is the volume of the cloud. The energy density can be expressed as a function of the cloud mass: $w_{\mathrm{TeV}}\approx3.8\times10^{-42}\times(M_{\mathrm{cloud}}/\mathrm{M}_{\odot})^{-1} \times L_{\gamma}(0.2 - 40\,\mathrm{TeV}) \approx 1.7 \times 10^3 \times (M_{\mathrm{cloud}}/\mathrm{M}_{\odot})^{-1} \times (d/\mathrm{kpc})^2\:\mathrm{eV}\,\mathrm{cm}^{-3}$. Assuming a mass of 7.2$\times 10^3$~M$_{\odot}$ and a distance of 11.3~kpc, estimated by Reynoso \& Mangum (2000), the energy density is estimated to $w_{\mathrm{TeV}} \approx\,30\:\mathrm{eV}\,\mathrm{cm}^{-3}$.

The proton energy distribution is assumed to follow a power-law with the same spectral index as the VHE $\gamma$-ray spectrum in the corresponding range and to lower energies. This is supported by the possible association with the EGRET source. The proton energy distribution can be extrapolated down to 1~GeV. The total proton energy above 1~GeV is estimated to $w_{>1\mathrm{GeV}}\, \approx\, 380 \:\mathrm{eV}\,\mathrm{cm}^{-3}$. Assuming that this density is uniform in a volume of radius $\sim5'$ as defined by the bright radio shell, this corresponds to a conversion efficiency of mechanical energy of the blast into CRs of $\eta_{\mathrm{CR}}\sim 0.3\times d_{11.3}^{5} \times M_{7.2}^{-1} \times E_{51}^{-1}$, where $d_{11.3} = d_{\mathrm{SNR}}/(11.3 \mathrm{kpc})$, $M_{7.2} = M_{\mathrm{cloud}}/(7.2\times10^3 \mathrm{M}_{\odot})$ and $E_{51} = E_{\mathrm{SNR}}/(10^{51} \mathrm{erg})$.

The above calculation has been made under the assumption that the whole VHE flux is emitted by the central cloud. The extension of the VHE gamma-ray source, however, indicates that at least parts of the neighboring clouds should also be involved in the emission. Including the three components likely associated with G348.5+0.1 into the calculation, the total mass accounts for 6.7$\times10^{4}$~M$_{\odot}$ and the lower limit on the conversion efficiency would be at the level of 4\%. The efficiency obtained with the central cloud can be considered as an upper limit. The estimated range of 4\% to 30\% for $\eta_{\mathrm{CR}}$ appears to be in good agreement with theoretical expectations. However, it should be noted that the cosmic ray density in the clouds may not be representative of the whole remnant. The interaction of the blast wave with the molecular clouds may affect the particle acceleration efficiency as well as the accelerated particle distribution in that region.

In the outlined hadronic scenario, secondary electrons might lead to significant X-ray synchrotron emission, especially in the high magnetic fields as suggested by the OH masers. The upper limit on the non-thermal X-ray emission from this region was only derived for the central cloud and is therefore underestimating the limit for the entire emission volume. Nevertheless, the non-detection of non-thermal X-ray emission from the molecular clouds is noteworthy and suggests that more detailed modeling is required for a more definitive conclusion.

\subsection{Leptonic scenario}

The most likely candidate for a leptonic origin of the VHE $\gamma$-ray emission is the plausible PWN seen in X-rays. In this case, the X-ray emission would be synchrotron emission of relativistic electrons which would radiate in the TeV range through the inverse-Compton (IC) process on an ambient radiation field. Assuming that the CMB is the main component of this target radiation field, the magnetic field at the nebula location can be constrained, according to $w_{\gamma}/w_{\mathrm{X}} \approx 0.1 (B/10\,\mu\mathrm{G})^{-2}$, where  $w_{\gamma}$ is the $\gamma$-ray integrated energy flux between 1 and 10~TeV and $w_{\mathrm{X}}$ the X-ray integrated energy flux between 0.5 and 10~keV (Aharonian et al. 1997). Assuming that the whole X-ray emission is contained in a region of extension $3^{\prime}20^{\prime \prime}\times1^{\prime}20^{\prime \prime}$ (corresponding to 95\% containment area of the X-ray PWN as estimated from the Chandra data), the ratio between the X-ray nebula extension and the H.E.S.S. source extension indicates that roughly a maximum of 7\% of VHE $\gamma$-rays could come from the X-ray nebula volume. The corresponding lower limit on the magnetic field in the X-ray nebula region is $\sim10\,\mu$G. This value is slightly larger than typical Galactic magnetic fields, and is not untypical for PWNe observed with H.E.S.S. (e.g. MSH~15-5\textit{2}, Aharonian et al. 2005).

According to Possenti et al. (2002), the pulsar spin-down luminosity can be estimated through the luminosity of the X-ray nebula. Assuming that the PWN is associated with the remnant and located at 11.3 kpc (Reynoso \& Mangum 2000), the X-ray luminosity $L_{\text{X} (2-10 \text{keV})} = 4.6\times10^{34}$~erg$\,$s$^{-1}$ implies a spin-down luminosity of $L_{\text{SD}} = 1.9\times10^{37}$~erg$\,$s$^{-1}$, with a lower limit of 2.8$\times10^{35}$~erg$\,$s$^{-1}$. Assuming the same distance for the VHE $\gamma$-ray source, the $\gamma$-ray luminosity between 1 and 10~TeV is $L_{\gamma (1-10 \text{TeV})} = 3.24\times10^{34}$~erg$\,$s$^{-1}$. It could easily be explained by IC emission from relativistic electrons accelerated within the PWN. A conversion efficiency of spin-down luminosity into VHE $\gamma$-rays of order 0.1\% is observed from several PWNe (e.g. Gallant 2007 and references therein). Deeper observations would be needed to confirm the PWN nature of the X-ray emission and to search for a pulsar at this location.

The $\gamma$-ray emission could also be produced by relativistic electrons accelerated by the blast-wave of the remnant. The non-detection of X-rays toward the molecular clouds is an argument against such a scenario. Due to the large magnetic field measured within the cloud, the VHE $\gamma$-ray luminosity between 1 and 10~TeV should be at least 10$^3$ times lower than the X-ray luminosity in the range 1~-~10~keV in a mean radiation field of $\sim$~eV$\,$cm$^{-3}$. The upper limit on the X-ray luminosity derived at the central cloud position would imply an energy density of the IC target radiation field higher than $10^3$~eV$\,$cm$^{-3}$ to explain the VHE $\gamma$-ray luminosity in a magnetic field of 100~$\mu$G. This estimate has been made assuming that this IC emission is dominantly from inside the clouds. If the emission comes mostly from electrons accelerated in lower density areas, the corresponding lower magnetic field would reduce the estimated IC target energy density down to more acceptable values. The lack of non-thermal X-rays from the shell would still require a higher IC target field than the CMB alone, even in a magnetic field as low as a few $\mu\mathrm{G}$, if one assumes that particles are still accelerated in the shock to very high energies.

The large gas density in the cloud could boost Bremsstrahlung from high energy electrons to detectable VHE $\gamma$-rays flux levels. The energy distribution of the produced $\gamma$-rays should then follow the same shape as that of the electrons. The large magnetic field measured at the cloud location implies that the electron distribution undergoes severe radiative losses at high energies. Assuming an age of, e.g., 2000~years for the remnant, the synchrotron loss time with a magnetic field of 0.5~mG indicates that above $\sim$30~GeV, the energy distribution should be steepened by one power, i.e. the spectral index  should be increased by one. If the entire detected VHE flux from HESS~J1714-385 (with $\Gamma \sim 2.3$) was attributed to Bremsstrahlung, this would point to an initial energy distribution of accelerated electrons with a very hard power law (spectral index of 1.3), which seems unlikely.

\section{Discussion}

A new VHE $\gamma$-ray source, HESS J1714-385, was discovered in positional coincidence with SNR G348.5+0.1. While the $\gamma$-rays are very likely emitted in processes associated with the SNR, the broadband data reveal a complex picture and allow different scenarios to interpret the $\gamma$-ray emission.

The size comparison of the VHE $\gamma$-ray source and the radio SNR, consisting of a partially well defined shell and an ``outbreak'', indicates that an association of the $\gamma$-ray source with the entire SNR shell is in principle possible. There are, however, strong indications that the SNR blast wave is interacting with several molecular clouds, as derived from CO emission and the presence of co-spatial OH masers. The detection of thermal X-ray emission from inside the NE remnant shell, which dominates the X-ray emission from G348.5+0.1, may provide further evidence that the SNR blast wave has interacted with dense molecular clouds; the thermal X-ray emission could be induced by similar processes as the ones seen in mixed-morphology SNRs. A natural scenario is then to attribute the detected $\gamma$-rays to hadronic processes predominantly taking place inside the shocked molecular clouds only. CR energetics derived under this assumption are compatible with standard efficiencies of CR acceleration in SNRs. This scenario also provides a natural explanation for the GeV emission (3EG J1714-3857) detected with EGRET in this region. However, the lack of non-thermal X-ray emission from the molecular cloud regions might pose a challenge to this scenario. Significant X-ray emission from secondary electrons, produced in interactions of hadronic cosmic rays with cloud gas, could be expected in the high magnetic fields indicated by the OH masers.

The VHE spectrum could also be explained by leptonic processes, either by IC scattering or Bremsstrahlung of high energy electrons. The absence of non-thermal X-ray synchrotron emission, as derived from the central molecular cloud position, renders such a leptonic scenario unlikely, if it can indeed be assumed that the VHE $\gamma$-ray emission is
emitted from inside the clouds; a hadron-dominated scenario would be more likely. However, since the VHE angular resolution is of the same order as the molecular cloud extensions, scenarios where high energy electrons are predominantly accelerated outside the clouds cannot be excluded. The lack of non-thermal X-ray emission from the radio rim disfavors a scenario of ongoing very high energy electron acceleration there. Another possible source of high energy electrons could be the X-ray PWN candidate \cxou, a non-thermal extended X-ray emission region seen in projection towards the NW of G348.5+0.1. The centroid of the H.E.S.S. source is not fully coincident with the X-ray source, but such offsets have already been observed in other $\gamma$-ray emitting PWN. The estimated spin-down luminosity of the potential pulsar powering the nebula as well as the conversion efficiency into $\gamma$-rays implied by the X-ray data appear reasonable. Given the number of associations of VHE $\gamma$-ray sources with PWNe, an association of HESS J1714-385 with the X-ray nebula seems plausible. There are, however, some remaining questions concerning the  identification of the nature of the non-thermal X-ray source. More detailed studies at other wavelengths will be helpful to confirm or reject the PWN nature of \cxou, in particular the search for a possible pulsar powering the nebula.

\begin{acknowledgements}
The support of the Namibian authorities and of the University of Namibia
in facilitating the construction and operation of H.E.S.S. is gratefully
acknowledged, as is the support by the German Ministry for Education and
Research (BMBF), the Max Planck Society, the French Ministry for Research,
the CNRS-IN2P3 and the Astroparticle Interdisciplinary Programme of the
CNRS, the U.K. Science and Technology Facilities Council (STFC),
the IPNP of the Charles University, the Polish Ministry of Science and 
Higher Education, the South African Department of
Science and Technology and National Research Foundation, and by the
University of Namibia. We appreciate the excellent work of the technical
support staff in Berlin, Durham, Hamburg, Heidelberg, Palaiseau, Paris,
Saclay, and in Namibia in the construction and operation of the
equipment. We wish to thank Estela Reynoso and Jeffrey Mangum for providing us their CO maps. 
\end{acknowledgements}

\end{document}